\documentclass[aps,pre,10pt,showpacs,floatfix,onecolumn,nofootinbib,a4paper]{revtex4-2}
\usepackage{amssymb, amsmath, amsthm, mathtools}
\usepackage{bm}
\usepackage{mathrsfs}
\usepackage{hyperref,bookmark}
\usepackage{graphicx}
\usepackage{epsfig,color}
\usepackage{mathrsfs}
\usepackage{verbatim}
\usepackage{url}
\usepackage{subcaption}
\usepackage{float}
\usepackage[greek,english]{babel}
\usepackage{dcolumn}

\newcommand{\tr}{\operatorname{tr}}
\newcommand{\dd}{\operatorname{d}\!}

\newcommand{\diver}{\operatorname{div}}
\newcommand{\curl}{\operatorname{curl}}

\newcommand{\n}{\bm{n}}
\newcommand{\e}{\bm{e}}

\newcommand{\free}{\mathscr{F}}

\newcommand{\vt}{\vartheta}

\newcommand{\WOF}{W_\mathrm{OF}}
\newcommand{\WQT}{W_\mathrm{QT}}
\newcommand{\el}{\xi_\mathrm{e}}
\newcommand{\bend}{\bm{b}}

\newcommand{\frameca}{(\e_x,\e_y,\e_z)}

\newcommand{\nigh}[1]{{\color{black}{#1}}}

\begin{document}
\latintext

\title{What  a twist cell experiment tells about a quartic twist theory for chromonics}
\author{Silvia Paparini}
\email{silvia.paparini@unipv.it}
\author{Epifanio G. Virga}
\email{eg.virga@unipv.it}
\affiliation{Department of Mathematics, University of Pavia, Via Ferrata 5, 27100 Pavia, Italy}
\begin{abstract}
The elastic theory of chromonic liquid crystals is not completely established. We know, for example, that for anomalously low twist constants (needed for chromonics) the classical Oseen-Frank theory may entail paradoxical consequences when applied to describe the equilibrium shapes of droplets surrounded by an isotropic phase: contrary to experimental evidence, they are predicted to dissolve in a plethora of unstable smaller droplets. We proposed a \emph{quartic twist} theory that prevents such an instability from happening. Here, we apply this theory to the data of \nigh{two} experiments devised to measure the planar anchoring strength at the plates bounding a twist cell filled with a chromonic liquid crystal; these data had before  been interpreted within the Oseen-Frank theory. We show that the quartic twist theory affords a better agreement with the experimental data, while delivering \nigh{in one case} a larger value for the anchoring strength.	
\end{abstract}
\date{\today}

\maketitle

\section{Introduction}\label{sec:intro}
Chromonic liquid crystals  are lyotropic phases, \nigh{here abbreviated LCLCs}, in which plank-shaped molecules arrange themselves in stacks when dissolved in solution (usually aqueous). For sufficiently large concentrations, the constituting stacks give rise to an ordered phase, either nematic or columnar \cite{lydon:chromonic_1998,lydon:handbook,lydon:chromonic_2010,lydon:chromonic,dierking:novel}.

Here, we shall only be concerned with the nematic phase. Numerous substances have a LCLC phase; these include organic dyes (especially those common in food industry), drugs, and oligonucleotides. Since LCLCs are mostly dispersed in water, they are also likely to have plenty of applications in life sciences \cite{shiyanovskii:real-time,mushenheim:dynamic,mushenheim:using,zhou:living}.

As pointed out in \cite{collings:anchoring}, successful applications of LCLCs presume the ability to control the orientation of the  director $\n$ in their nematic phase, and this can only be done by characterizing the anchoring performance of diverse surface treatments. Planar anchoring of LCLCs on surfaces has been achieved by a number of methods, which include irradiation with polarized light of azo-polymer substrates \cite{yip:photo-patterned,ichimura:surface-assisted,fujiwara:surface-assisted,van_der_Asdonk:patterning,peng:patterning}, mechanical rubbing of polymer surface coatings \cite{nastishin:optical,tone:dynamical} as well as of bare glass \cite{zhou:elasticity_2012, mcginn:planar}, and fabrication of closely spaced surface ridges \cite{yi:orientation,kim:macroscopic}.

The alignment method adopted in \cite{collings:anchoring}, one of the experimental works that motivated our theoretical investigation, uses commercially available rubbed polyimide (PI) layers, known to induce a good planar alignment on solutions of disodium cromoglycate (DSCG), an anti-asthmatic drug.

When a nematic liquid crystal is placed in a cell between two parallel plates promoting planar (weak) anchoring along directions at right angles to one another, the nematic director $\n$ acquires a twist distortion in the cell. If the anchoring strength $W_0$ is not infinite, however, $\n$ at the plates departs by an \emph{offset} angle $\delta$ from the preferred \emph{easy} axis, so that the overall twist angle $\Omega$ is less than $\pi/2$. Clearly, the smaller is $W_0$, the larger is $\delta$.

Building on earlier work \cite{mcginn:planar}, the experiment performed in \cite{collings:anchoring} determined $W_0$ at the plates bounding a cell filled with a DSCG aqueous solution by measuring optically $\Omega$. The measured value of $W_0$ turned out to be less than $1\,\mu\mathrm{J}/\mathrm{m}^2$. \nigh{A larger value of $W_0$ was measured in \cite{peng:patterning}, which adopted an experimental method similar to that used in \cite{collings:anchoring}, but with a different substrate.}

The elastic theory employed in \nigh{both \cite{collings:anchoring} and \cite{peng:patterning}} is the classical quadratic Oseen-Frank theory, with an anomalously small twist constant $K_{22}$. In particular, $K_{22}$ must be smaller than the saddle-splay constant $K_{24}$, thus violating one of the inequalities Ericksen~\cite{ericksen:inequalities} had put forward to guarantee that the Oseen-Frank stored energy be bounded below. However, as shown in \cite{paparini:paradoxes}, free-boundary problems may reveal noxious consequences stemming from violating Ericksen's inequalities.   If $K_{22}<K_{24}$, a LCLC droplet surrounded by an isotropic fluid environment (enforcing degenerate planar anchoring for the director at the interface) is predicted to be unstable against \emph{shape} perturbations: it would split indefinitely in smaller droplets, while the total free energy plummets to negative infinity \cite{paparini:paradoxes}.

This prediction is in sharp contrast with the wealth of experimental observations of LCLC tactoidal\footnote{\emph{Tactoids} are elongated, cylindrically symmetric shapes with pointed ends as poles.} droplets in the biphasic region of phase space, where nematic and isotropic phases coexist in equilibrium \cite{tortora:self-assembly,tortora:chiral,peng:chirality,nayani:using,shadpour:amplification}. These studies have consistently reported stable twisted bipolar tactoids.

To resolve this contradiction, in \cite{paparini:elastic} we proposed a minimalist quartic theory for LCLCs, which adds to the Oseen-Frank energy density a single quartic term in the twist measure; hence the name \emph{quartic twist} theory. We showed in \cite{paparini:elastic} that indeed within this theory the total free energy of chromonic droplets subject to degenerate planar interfacial anchoring remains bounded below, even if $K_{22}<K_{24}$.   

In \cite{paparini:spiralling}, the quartic twist theory was applied to explain the formation of \emph{inversion rings} within spherical cavities enclosing water solutions of SSY (Sunset Yellow, a dye used in industrially processed food) subject to \emph{homeotropic} boundary conditions for $\n$. The predictions of the theory were recently corroborated by some experimental evidence \cite{spina:intercalation,ciuchi:inversion}.

Here, we reexamine the experiments performed in \cite{collings:anchoring} and \cite{peng:patterning} in the light of the quartic twist theory. We show that this also predicts a \emph{linear} twist between the cell's plates with an offset angle related to both the anchoring strength $W_0$ and a phenomenological length $a$ featuring in the theory. The data presented in both \cite{collings:anchoring} and \cite{peng:patterning} turn out to be better fitted by this theory compared to the classical quadratic one (\nigh{with errors decreased by $3\,\%$ and $1.4\,\%$, respectively});  as a result, both $W_0$ and $a$ are determined,  the former with a value larger than that found in \cite{collings:anchoring} \nigh{and much closer to that found in \cite{peng:patterning}}.

The rest of the paper is organized as follows. In Section~\ref{sec:theory}, we summarize the quartic twist theory employed here, recalling only its essential features. In Section~\ref{sec:experiment}, we apply this theory to a $\pi/2$ twist cell with weak planar anchoring and we employ it to reinterpret the experimental data of \cite{collings:anchoring} and \cite{peng:patterning}. Finally, we summarize our conclusions in Section~\ref{sec:conclusion}. The paper is closed by two appendices: one shows that the weak anchoring is never fully broken in the case under study, and the other gives details on the error estimates involved in data fitting. 

\section{Quartic Twist Theory}\label{sec:theory}
What makes chromonic nematics differ from ordinary ones is the \emph{ground state} of their distortion: a \emph{double} twist for the former, a uniform field (along any selected direction) for the latter. We now explore this difference in more detail.

We follow Selinger~\cite{selinger:interpretation} in writing the elastic energy density of the Oseen-Frank theory in an equivalent form,
\begin{equation}
	\label{eq:Frank_equivalent}
	\WOF(\n,\nabla\n)=\frac12(K_{11}-K_{24})S^2+\frac12(K_{22}-K_{24})T^2+\frac12K_{33}B^2+2K_{24}q^2,
\end{equation}
where $S:=\diver\n$ is the \emph{splay}, $T:=\n\cdot\curl\n$ is the \emph{twist}, $B^2:=\bend\cdot\bend$ is the square modulus of the \emph{bend} vector $\bend:=\n\times\curl\n$, and $q>0$ is the \emph{octupolar splay}  \cite{pedrini:liquid} derived from the following equation
\begin{equation}
	\label{eq:identity}
	2q^2=\tr(\nabla\n)^2+\frac12T^2-\frac12S^2.
\end{equation}
Since $(S,T,B,q)$ are independent \emph{distortion measures}, it easily follows from \eqref{eq:Frank_equivalent} that $\WOF$ is positive semi-definite whenever
\begin{subequations}\label{eq:Ericksen_inequalities}
	\begin{eqnarray}
		&K_{11}\geqq K_{24}\geqq0,\label{eq:Ericksen_inequalities_1}\\
		&K_{22}\geqq K_{24}\geqq0, \label{eq:Ericksen_inequalities_2}\\
		&K_{33} \geqq 0,\label{eq:Ericksen_inequalities_3}
	\end{eqnarray}
\end{subequations}
which are the celebrated \emph{Ericksen's inequalities} \cite{ericksen:inequalities}. If these inequalities are satisfied in strict form, the global ground state of $\WOF$ is attained on the uniform director field, characterized by
\begin{equation}
	\label{eq:uniform_ground_state}
	S=T=B=q=0,
\end{equation}
which designates the ground state of ordinary nematics.

LCLCs are characterized by a different ground state, one where all distortion measures vanish, \emph{but} $T$, which we call a \emph{double twist}. Here, we adopt the terminology proposed by Selinger~\cite{selinger:director} (see also \cite{long:explicit}) and distinguish between \emph{single} and \emph{double} twists. The former is characterized by
\begin{equation}
	\label{eq:single_twist}
	S=0,\quad B=0,\quad T=\pm2q,
\end{equation}
which designates a director distortion capable of filling \emph{uniformly} the whole space \cite{virga:uniform}.\footnote{It is nothing but what others would call a \emph{cholesteric} twist.} Unlike this, a double twist \emph{cannot} fill space uniformly: it can possibly be realized locally, but not everywhere. In words, we also say that it is a \emph{frustrated} ground state.\footnote{It was shown in \cite{paparini:stability} that  a double twist can, for example, be attained exactly on the symmetry axis of cylinders enforcing degenerate planar anchoring on their lateral boundary.} For the Oseen-Frank theory to accommodate such a ground state, inequality \eqref{eq:Ericksen_inequalities_2} must be replaced by $K_{24}\geqq K_{22}\geqq0$, but this comes at the price of making $\WOF$ unbounded below \cite{paparini:stability}.

The essential feature of the quartic twist theory proposed in \cite{paparini:elastic} is to envision a double twist (with two equivalent chiral variants) as ground state of LCLCs in three-dimensional space,
\begin{equation}
	\label{eq:double_twist}
	S = 0, \quad T = \pm T_0, \quad B = 0, \quad q = 0.
\end{equation}
The degeneracy of the ground  double twist  in \eqref{eq:double_twist} arises from  the achiral nature of the molecular aggregates that constitute these materials, which is reflected in the lack of chirality of their condensed phases.

The elastic stored energy must equally penalize both ground chiral variants. Our minimalist  proposal to achieve this goal was to add  a \emph{quartic twist} term to the Oseen-Frank stored-energy density, 
\begin{equation}
	\label{eq:quartic_free_energy_density}
	\WQT(\n,\nabla\n):=\frac{1}{2}(K_{11}-K_{24})S^2+\frac{1}{2}(K_{22}-K_{24})T^2+ \frac{1}{2}K_{23}B^{2}+\frac{1}{2}K_{24}(2q)^2 + \frac14K_{22}a^2T^4,
\end{equation}
where $a$ is a \emph{characteristic length}. $\WQT$ is bounded below whenever 
\begin{subequations}\label{eq:new_inequalities}
	\begin{eqnarray}
		&K_{11}\geqq K_{24}\geqq0,\label{eq:new_inequalities_1}\\
		&K_{24}\geqq K_{22}\geqq0, \label{eq:new_inequalities_2}\\
		&K_{33}\geqq0.\label{eq:new_inequalities_3}
	\end{eqnarray}
\end{subequations}
If these inequalities  hold, as we shall assume here, then $\WQT$ is minimum at the  degenerate double twist \eqref{eq:double_twist}
characterized by
\begin{equation}
	\label{eq:T_0min}
	T_0:=\frac{1}{a}\sqrt{\frac{K_{24}-K_{22}}{K_{22}}}.
\end{equation}
Here, we shall treat $a$ as a phenomenological parameter to be determined experimentally.

For an aligning substrate with easy axis along the unit vector $\e$, we write the anchoring energy density (per unit area) as
\begin{equation}
	\label{eq:anchoring_energy}
	W_\mathrm{a}:=\frac12W_0\left(1-(\n\cdot\e)^2\right),
\end{equation}
where $W_0>0$ is the \emph{anchoring strength} \cite{rapini:distortion}.

\begin{figure}[h] 
	\centering
	\begin{subfigure}[c]{0.4\linewidth}
	\includegraphics[width=\linewidth]{3d_fin_a.pdf}
	\caption{ The two bounding plates exhibit their easy axes as parallel (red) lines aligned to the unit vectors $\e_x$ and $\e_y$, respectively, of a Cartesian frame $\frameca$.}
	\label{fig:sketch_a}
	\end{subfigure}
\qquad\qquad
    \begin{subfigure}[c]{0.3\linewidth}
    	\includegraphics[width=\linewidth]{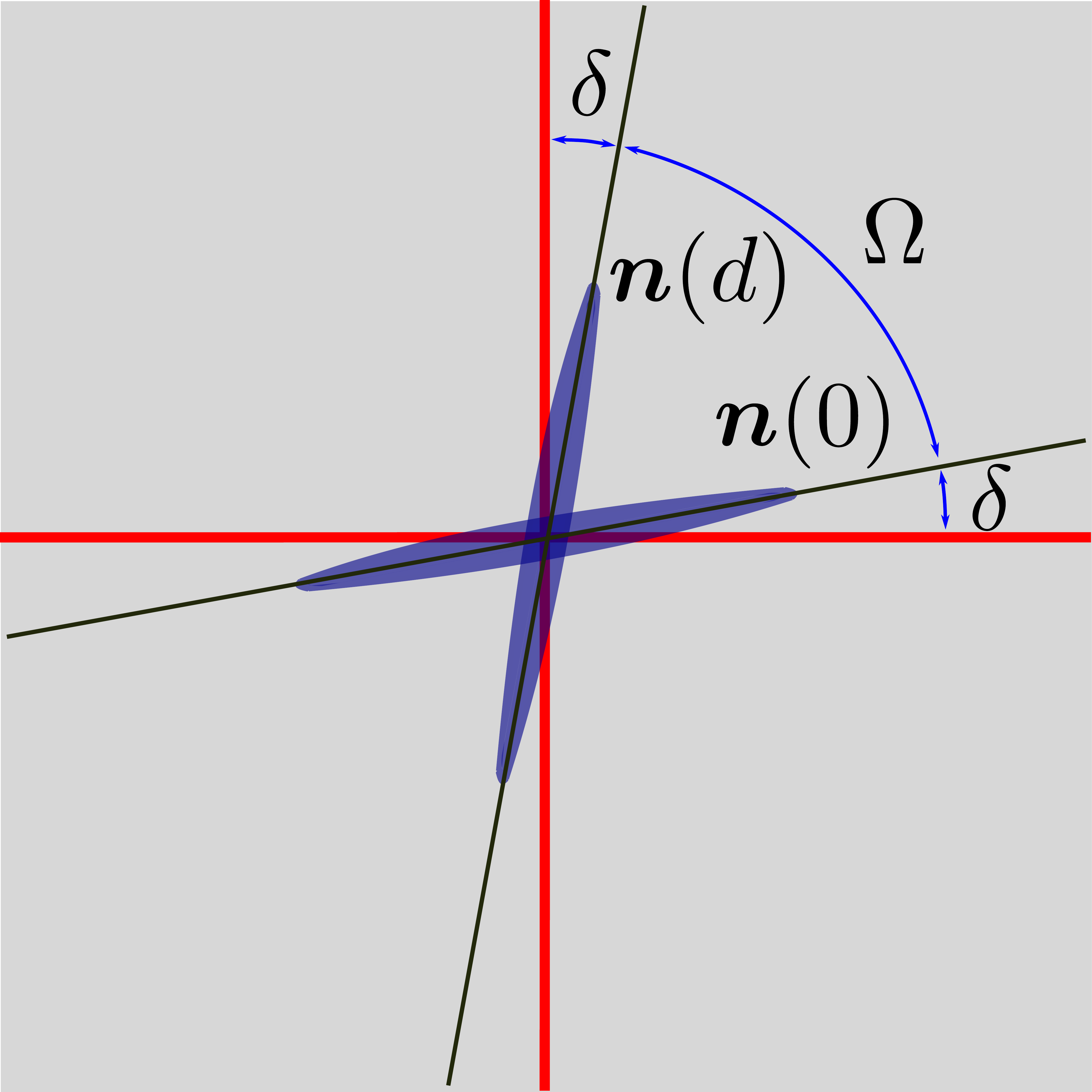}
    	\caption{Top view (along $\e_z$) illustrating both the offset angle $\delta$ and the total twist angle $\Omega$.}
    	\label{fig:sketch_b}
    	\end{subfigure}
	\caption{Schematic of a $\pi/2$ twist cell.}
	\label{fig:sketch}
\end{figure}

In a $\pi/2$ twist cell bounded by two plates with area $A$ at a distance $d$, both parallel to the $(x,y)$ plane of a Cartesian frame $\frameca$, we take $\e=\e_x$ for the plate at $z=0$ and $\e=\e_y$ for the plate at $z=d$ (see Fig.~\ref{fig:sketch}).
Moreover, we assume that the nematic director within the cell is given the form
\begin{equation}
	\label{eq:n_twist}
	\n(z)=\cos\vt(z)\e_x+\sin\vt(z)\e_y,
\end{equation}
where the twist angle $\vt$ is assumed to be a smooth function of $z$ only. It readily follows from \eqref{eq:n_twist} that
\begin{equation}
	\label{eq:nabla_n}
	\nabla\n=\vt'\left(\cos\vt\e_y\otimes\e_z-\sin\vt\e_x\otimes\e_z\right),
\end{equation}
which, also with the aid of \eqref{eq:identity}, implies in turn that 
\begin{equation}
	\label{eq:twist_measures}
	S=0,\quad T=-\vt',\quad B=0,\quad2q^2=\frac12T^2.
\end{equation}

By combining \eqref{eq:quartic_free_energy_density} and \eqref{eq:anchoring_energy}, we write the total energy stored in the cell as the functional
\begin{equation}
	\label{eq:total_energy_functional}
	\free[\vt]:=\frac12A\bigg\{\int_0^dK_{22}\left(\vt'^2+\frac12a^2\vt'^4\right)\dd z-W_0\left(\cos^2\vt(0)+\sin^2\vt(d)\right)\bigg\},
\end{equation}
where a prime $'$ denotes differentiation with respect to $z$ and an inessential additive constant ($AW_0$) has been omitted. Letting $\zeta:=z/d$, we can write $\free$ in the following dimensionless form
\begin{equation}
	\label{eq:total_energy_dimensionless}
	\mathcal{F}[\vt]:=\frac{d\free[\vt]}{AK_{22}}=\frac12\bigg\{\int_0^1\left(\vt'^2+\frac12\lambda^2\vt'^4\right)\dd\zeta-\alpha\left(\cos^2\vt(0)+\sin^2\vt(1)\right)\bigg\},
\end{equation}
where $\lambda$ and $\alpha$ are dimensionless parameters defined as 
\begin{equation}
	\label{eq:lambda_alpha}
	\lambda:=\frac{a}{d},\quad\text{and}\quad\alpha:=\frac{d}{\el},\quad\text{with}\quad\el:=\frac{K_{22}}{W_0}.
\end{equation}
The parameter $\alpha$  is a relative measure of anchoring strength: the anchoring is \emph{strong} if the \emph{extrapolation} length $\el$ is small compared to $d$, it is \emph{weak} if $\el$ is large compared with $d$.   

The Euler-Lagrange equations associated with $\mathcal{F}$ is
\begin{equation}
	\label{eq:E_L_equation}
	(\vt+\lambda^2\vt'^3)'=0
\end{equation}
subject to the boundary conditions
\begin{subequations}\label{eq:boundary_conditions}
	\begin{eqnarray}
	&(\vt'+\lambda^2\vt'^3)|_{\zeta=0}=\frac12\alpha\sin2\vt(0),\label{eq:boundary_conditions_0}\\
	&(\vt'+\lambda^2\vt'^3)|_{\zeta=1}=\frac12\alpha\sin2\vt(1).\label{eq:boundary_conditions_1}
\end{eqnarray}
\end{subequations}
Equations \eqref{eq:E_L_equation} and \eqref{eq:boundary_conditions} require that there is a constant $c$ such that 
\begin{equation}
	\label{eq:E_L_reduced}
	\vt'+\lambda^2\vt'^2=c
\end{equation}
and
\begin{equation}
	\label{eq:boundary_conditions_reduced}
	c=\frac{\alpha}{2}\sin2\vt(0)=\frac{\alpha}{2}\sin2\vt(1).
\end{equation}
It follows from \eqref{eq:E_L_reduced} that also $\vt'$ is constant throughout the cell, and from \eqref{eq:boundary_conditions_reduced} that there are two classes of equilibrium solutions, characterized by the conditions 
\begin{subequations}\label{eq:equilibria}
	\begin{eqnarray}
&\vt(1)=\vt(0)+k\pi,\label{eq:equilibrium_1}\\
&\vt(1)=\frac{\pi}{2}-\vt(0)+k\pi,\label{eq:equilibrium_2}
\end{eqnarray}
\end{subequations}
where $k\in\mathbb{Z}$. Since
\begin{equation}
	\label{eq:theta_prime}
	\vt'\equiv\vt(1)-\vt(0),
\end{equation}
the less distorted (and less energetic) representatives of both classes of solutions \eqref{eq:equilibria} are obtained for $k=0$.

Solution \eqref{eq:equilibrium_1} thus corresponds to the case of \emph{broken} anchoring, for which $c=0$, $\vt\equiv\vt_0$, and
\begin{eqnarray}
	\label{eq:energy_c_0}
	\mathcal{F}[\vt_0]=-\frac{\alpha}{2}\quad\forall\ \vt_0.
\end{eqnarray}
Letting $\delta:=\vt(0)$, solution \eqref{eq:equilibrium_2}, once combined with \eqref{eq:boundary_conditions_reduced} and \eqref{eq:theta_prime}, demands that 
\begin{equation}
	\label{eq:g_equation}
	\vt'=\frac{\pi}{2}-2\delta=g_{\lambda}(\alpha,\delta),
\end{equation}
where
\begin{equation}
	\label{eq:g_definition}
	g_\lambda(\alpha,\delta):=\frac{1}{\lambda}\frac{6^{2/3}\left(\sqrt{(9\alpha\lambda\sin2\delta)^2+48}+9\alpha\lambda\sin2\delta\right)^{2/3}-12}{6^{4/3}\left(\sqrt{(9\alpha\lambda\sin2\delta)^2+48}+9\alpha\lambda\sin2\delta\right)^{1/3}}
\end{equation}
is the real root of the cubic equation
\begin{equation}
	\label{eq:cubic_equation}
	g(1+\lambda^2g^2)=\frac{\alpha}{2}\sin2\delta.
\end{equation}
It is shown in Appendix~\ref{sec:comparison} that the total energy associated with the latter equilibrium solution is always less that $\mathcal{F}[\vt_0]$ in \eqref{eq:energy_c_0}, so that the anchoring is never broken.

\begin{figure}[h] 
	\centering
	\includegraphics[width=0.4\linewidth]{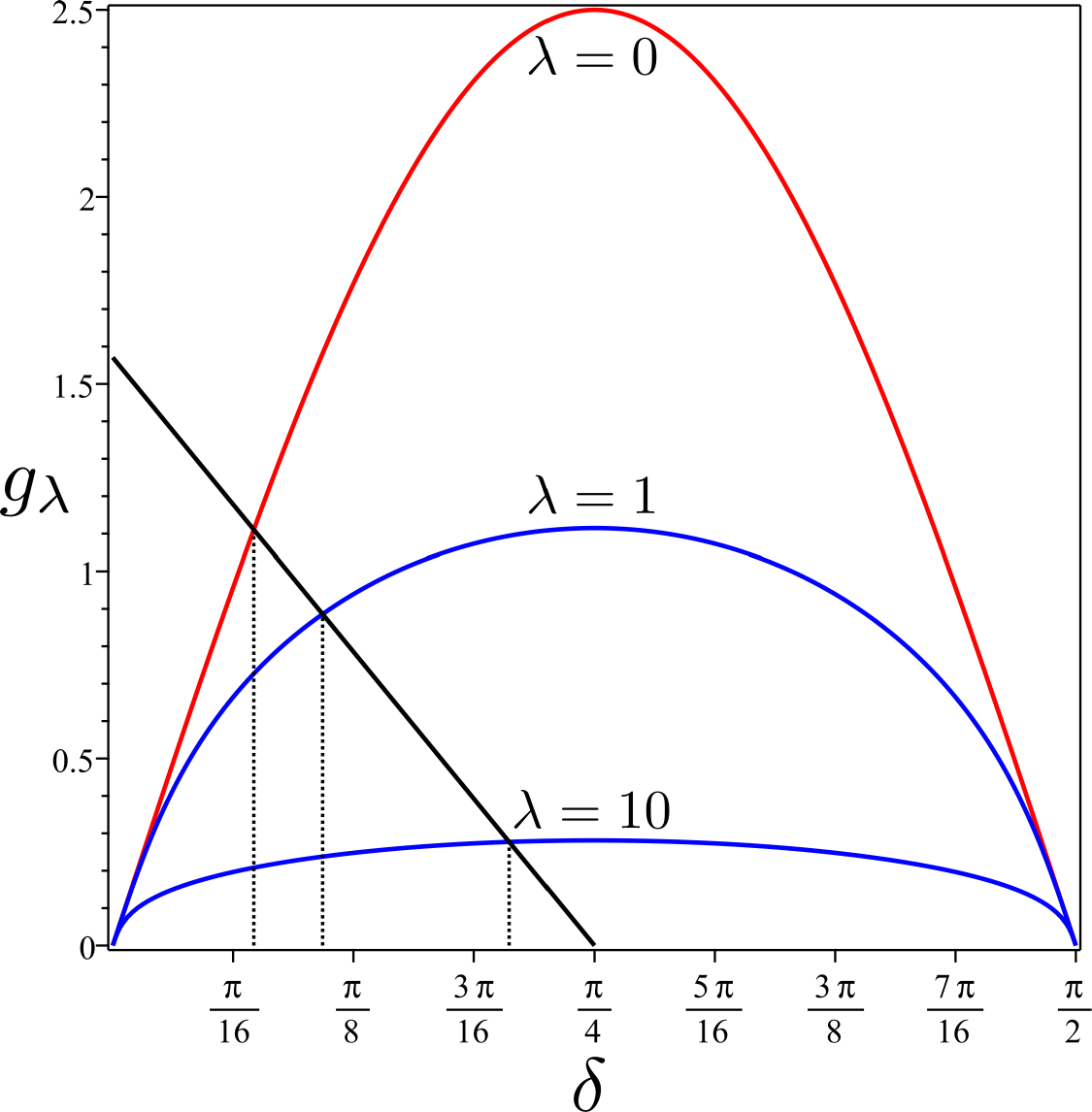}
	\caption{Graphical solution of \eqref{eq:g_equation}. The graph of $g_\lambda(\alpha,\delta)$ against $\delta$ is drawn for $\alpha=5$ and several values of $\lambda$. For $\lambda=0$, the red curve represents the graph of $g_0$ in \eqref{eq:g_0}. The black straight line is the graph of $\frac{\pi}{2}-2\delta$, which has a single intersection with the graph of each $g_\lambda$.}
	\label{fig:g_delta}
\end{figure}

\begin{figure}[h] 
	\begin{subfigure}[c]{0.4\linewidth}
	\centering
	\includegraphics[width=\linewidth]{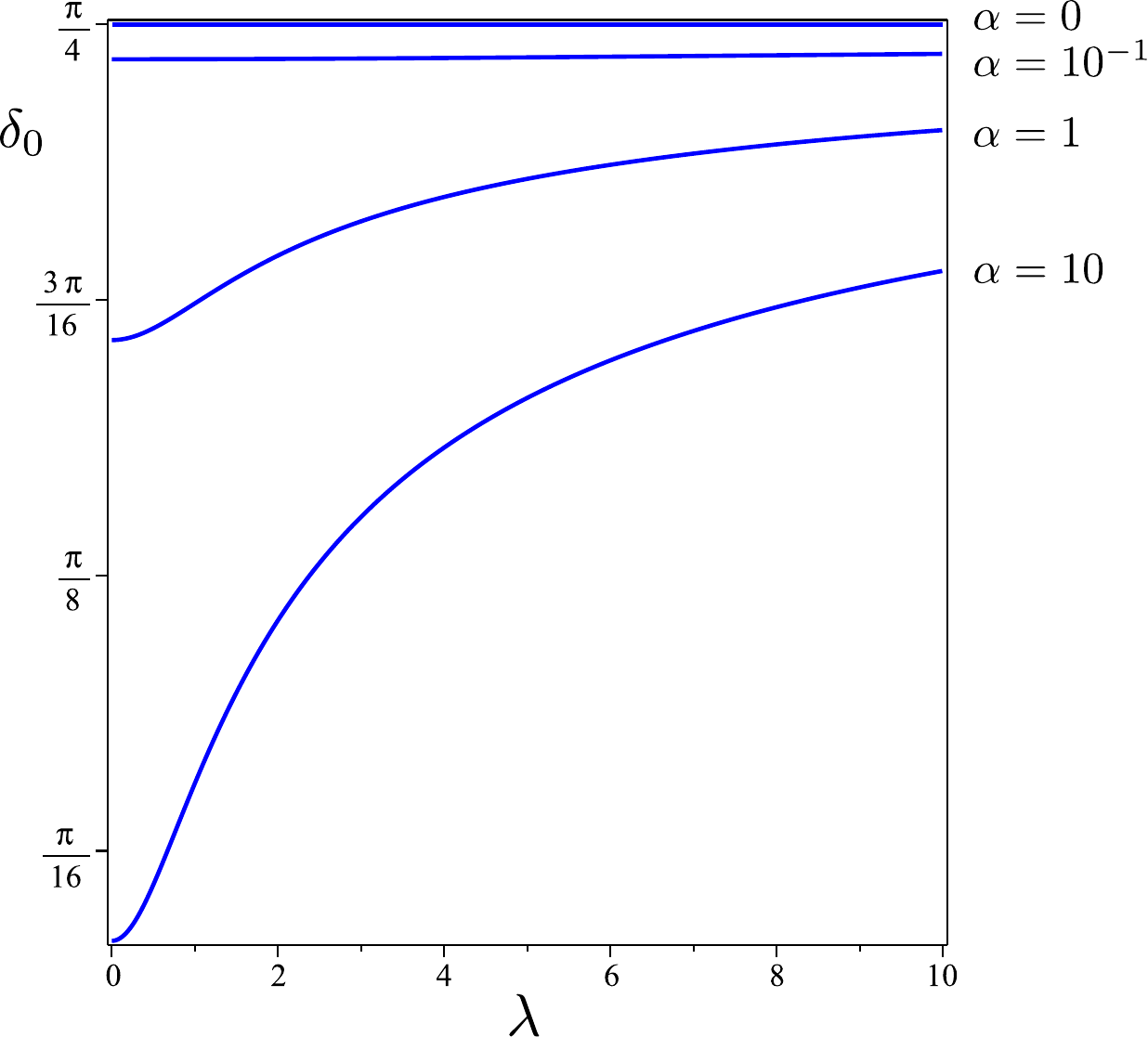}
	\caption{$\delta_0$ against $\lambda$ for several values of $\alpha$.}
	\label{fig:delta_0_lambda}
\end{subfigure}
\qquad
\begin{subfigure}[c]{0.4\linewidth}
	\centering
	\includegraphics[width=\linewidth]{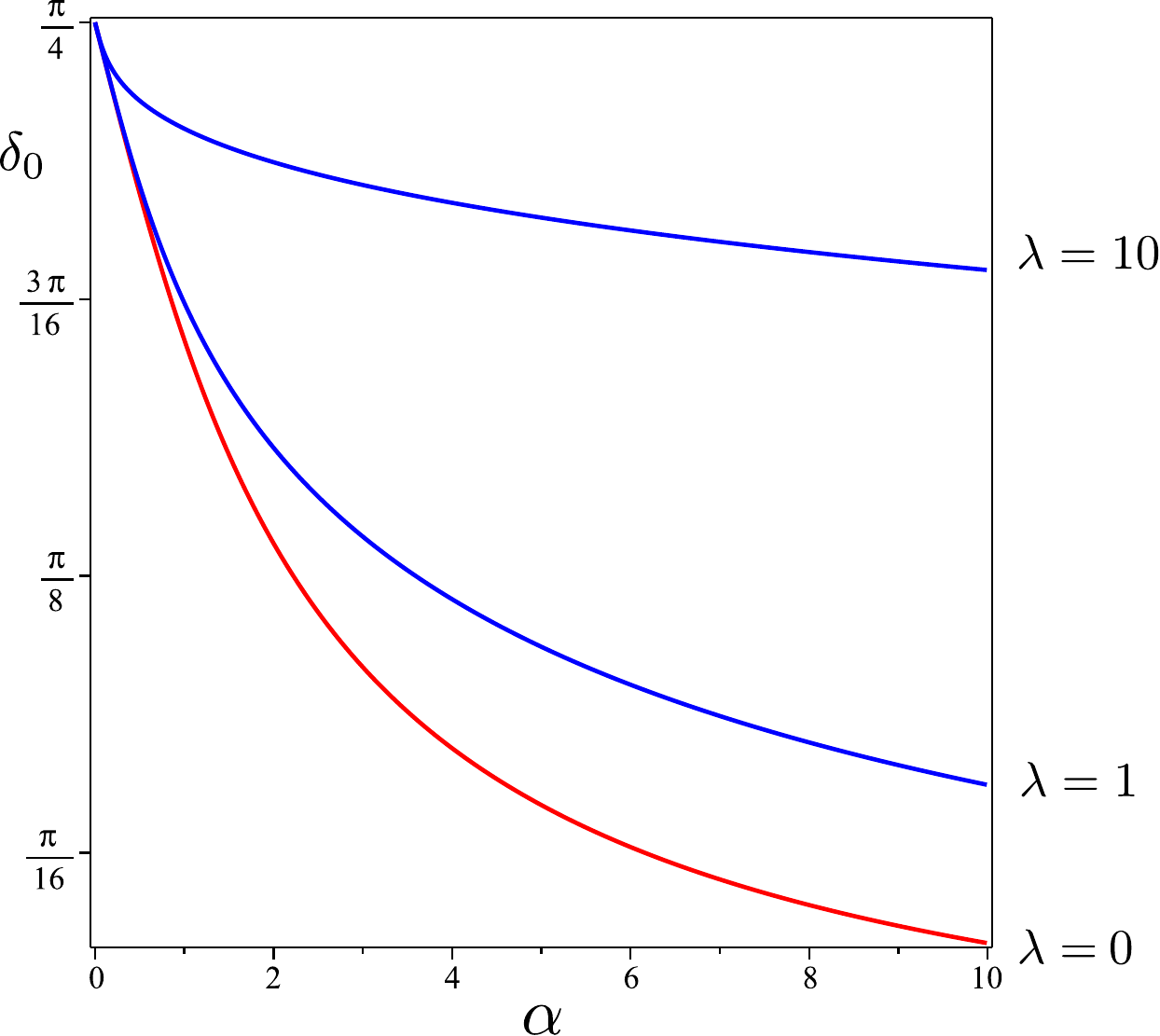}
	\caption{$\delta_0$ against $\alpha$ for several values of $\lambda$.}
	\label{fig:delta_0_alpha}
	\end{subfigure}
\caption{Plots showing how the root $\delta_0$ of \eqref{eq:g_equation} depends on both $\lambda$ and $\alpha$}
\label{fig:delta_plots}
\end{figure}

Figure~\ref{fig:g_delta}
illustrates the graphical solution of \eqref{eq:g_equation}: there is only one intersection between the graph of $g_\lambda$ and the straight line $\frac{\pi}{2}-2\delta$; it falls for $\delta=\delta_0(\alpha,\lambda)$.
Figure~\ref{fig:delta_plots} shows how $\delta_0$ depends on both $\lambda$ and $\alpha$.
In the limit as $\lambda$ tends to $0$, we retrieve the quadratic case already studied in \cite{collings:anchoring} and \cite{peng:patterning}, for which \eqref{eq:g_definition} reduces to
\begin{equation}
	\label{eq:g_0}
	g_0(\alpha,\delta):=\frac{\alpha}{2}\sin2\delta.
\end{equation}
For $\alpha$ small, $\delta_0(\alpha,\lambda)$ approaches $\frac{\pi}{4}$ for all values of $\lambda$. When $\alpha$ increases, $\delta_0(\alpha,\lambda)$ is an increasing function of $\lambda$ that exhibits a larger excursion as $\alpha$ increases. For any given value of $\lambda$, $\delta_0(\alpha,\lambda)$ tends to $\frac{\pi}{4}$ as $\alpha$ tends to $0$ and decreases monotonically as $\alpha$ increases, with an excursion that grows larger as $\lambda$ decreases (see Fig.~\ref{fig:delta_plots}).
The total equilibrium twist angle  is then given as a function of $(\alpha,\lambda)$ by 
\begin{equation}
	\label{eq:total_twist}
	\Omega(\alpha,\lambda)=\frac{\pi}{2}-2\delta_0(\alpha,\lambda).
\end{equation} 

We shall see in the following section how $\Omega$ has effectively been measured by optical means in \cite{collings:anchoring} and \cite{peng:patterning}. We shall fit their data to determine both $\alpha$ and $\lambda$ (that is, both $W_0$ and $a$, once $K_{22}$ is obtained from independent measurements).

\section{Experiments Reinterpretation}\label{sec:experiment}
\nigh{Here, we reexamine the experiments performed in \cite{collings:anchoring} and \cite{peng:patterning}.} 

In \cite{collings:anchoring}, an aqueous  solution of disodium cromoglycate (DSCG) (at a concentration $16.0\,\%\, \mathrm{(wt/wt)}$ and temperature $25^\circ\mathrm{C}$) was used to fill a $\pi/2$ twist cell bounded by  plates with different rubbed PI aligning substrates. Linearly polarized monochromatic light was shone perpendicularly on one plate of the cell and detected on the other plate through a linear polarizer aligned in a direction either orthogonal or parallel to the polarization of the incoming light. The cell was  rotated around the direction of light propagation  and the transmitted intensity $I(\theta)$ (scaled to the intensity of the incoming light) was monitored as a function of the rotation angle $\theta$.

A theory put forward by McIntyre~\cite{mcintyre:light,mcintyre:transmission} related (in closed form) the total twist angle $\Omega$ to both the \nigh{minimum and maximum values of $I(\theta)$, $I_\perp^\mathrm{min}$ and $I_\perp^\mathrm{max}$, respectively,  when the polarizers at the bounding plates are orthogonal to one another. Following \cite{ong:origin} and \cite{mcginn:planar} (see also p.\,174 of \cite{khoo:liquid}), we give these scaled intensities the following expressions,
\begin{subequations}\label{eq:I_def}
\begin{align}
			I_{\perp}^{\mathrm{min}}(\Omega)&=\left(\cos\tau\sin\Omega-\frac{\Omega}{\tau}\sin\tau\cos\Omega\right)^2,\label{eq:I_min_perp}\\
			I_{\perp}^{\mathrm{max}}(\Omega)&=\left(1-\frac{\Omega^2}{\tau^2}\right)\sin^2\tau+\left(\cos\tau\sin\Omega-\frac{\Omega}{\tau}\sin\tau\cos\Omega\right)^2,			\label{eq:I_max_perp}
\end{align}
where $\tau:=\sqrt{4\Omega^2+\Psi^2}/2$ and $\Psi:=(2\pi/\lambda_0)d\Delta n$ is the phase retardation angle; here $\lambda_0$ is  the wavelength of the light in vacuum and $\Delta n$ is the birefringence of the material.\footnote{\nigh{Equation \eqref{eq:I_min_perp} is equivalent to equation (1) of \cite{peng:patterning}, provided in the latter one corrects the typo that made appear $(X\sin\tau/\sqrt{1+X^2})^2$  as $(\sin\tau/\sqrt{1+X^2})^2$. Moreover, a perfectly equivalent form of \eqref{eq:I_max_perp} would be the following,
\begin{equation*}
I_{\perp}^{\mathrm{max}}(\Omega)=\left(1-\frac{\Omega^2}{\tau^2}\right)\sin^2\Omega+\left(\cos\Omega\sin\tau-\frac{\Omega}{\tau}\sin\Omega\cos\tau\right)^2.
\end{equation*}}} Similarly, the minimum and maximum (scaled) intensities, $I_\parallel^\mathrm{min}$ and $I_\parallel^\mathrm{max}$, transmitted through parallel polarizers are given by
\begin{align}
	I_{\parallel}^{\mathrm{min}}(\Omega)&=1-I_{\perp}^{\mathrm{max}}(\Omega),\label{eq:I_min_para}\\
I_{\parallel}^{\mathrm{max}}(\Omega)&=1-I_{\perp}^{\mathrm{min}}(\Omega).\label{eq:I_max_para}
\end{align}
\end{subequations}
All intensities in \eqref{eq:I_def} are very sensitive to $\Omega$ \cite{mcginn:planar}. On the other hand,
as pointed out in \cite{collings:anchoring}, both these and the \emph{intensity ratio}
\begin{equation}
	\label{eq:R_def}
	R(\Omega):=\frac{I_{\perp}^\mathrm{max}(\Omega)-I_{\parallel}^\mathrm{min}(\Omega)}{I_{\perp}^\mathrm{max}(\Omega)+I_{\parallel}^\mathrm{min}(\Omega)}=2I_{\perp}^\mathrm{max}(\Omega)-1,
\end{equation}
which ranges in the interval $[-1,1]$, are rather insensitive to the incident light intensity.

In \cite{collings:anchoring}, $R$ was measured at different values of the thickness $d$ in a wedge cell with rubbed SE-7511L PI substrates for $\lambda_0=633\, \mathrm{nm}$ and  $\Delta n=-0.0184$ (measured 
in \cite{collings:anchoring} for DSCG at concentration $16.0\,\% (\mathrm{wt/wt})$ and temperature $25^\circ\mathrm{C}$).} 

The data taken from Fig.~4 of \cite{collings:anchoring} are represented by dots in  Fig.\ref{fig:R_vs_d} below; they can be fitted to the theoretical functions in \eqref{eq:R_def} and \eqref{eq:I_max_perp}, with the total twist angle $\Omega$ given by \eqref{eq:total_twist}, for both $\lambda=0$ and $\lambda>0$.
\begin{figure}[h] 
\centering
	\includegraphics[width=0.4\linewidth]{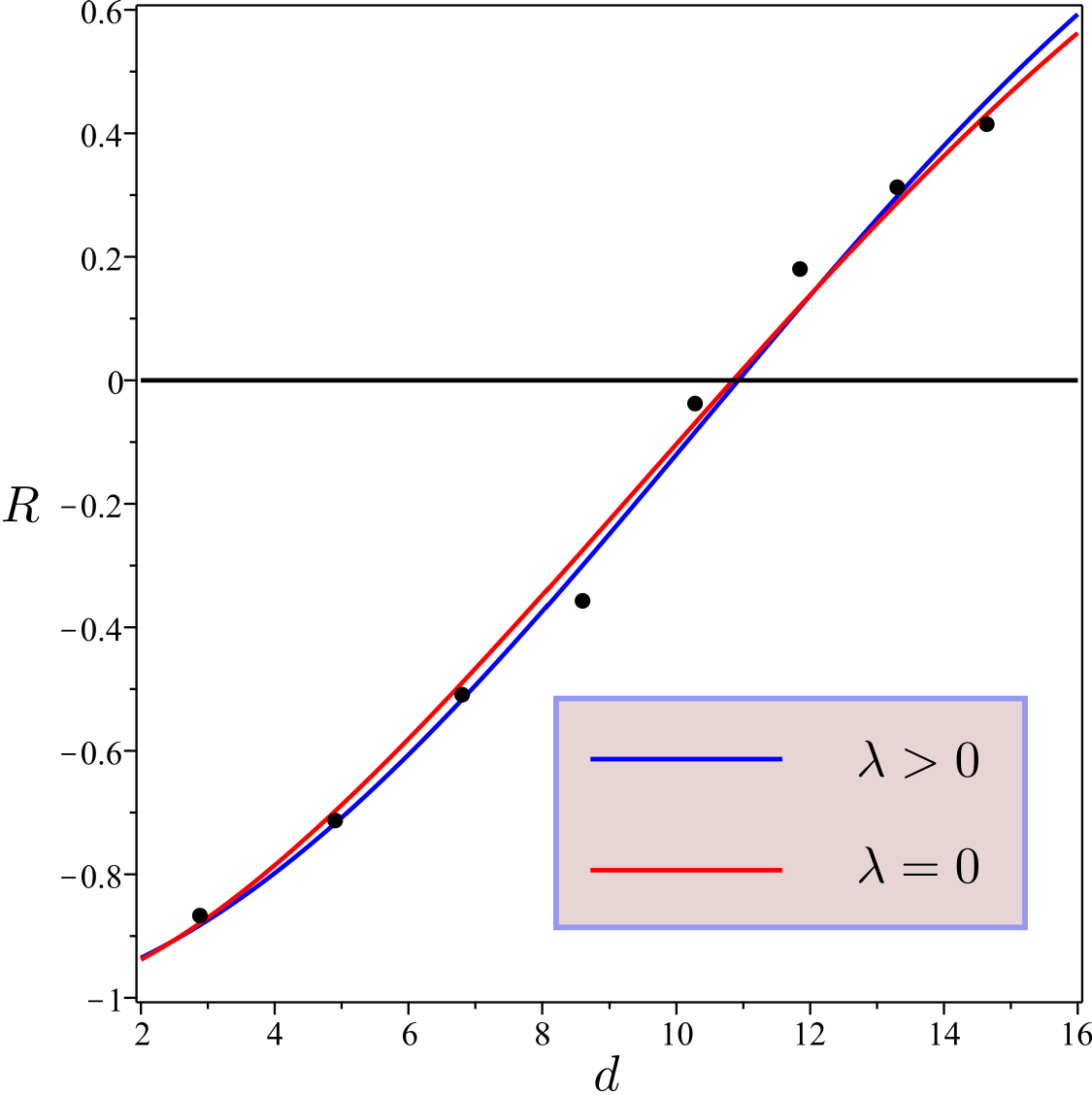}
\caption{Best fits of the data for the intensity ratio $R$  in Fig.~4 of \cite{collings:anchoring} with the theoretical functions given by \eqref{eq:R_def} and \eqref{eq:I_max_perp} when the  total twist angle $\Omega$ is delivered by \eqref{eq:total_twist}.  Red curve (classical quadratic theory): $\lambda=0$ and $\alpha$ is the only fitting parameter. Blue curve (quartic twist theory): $\lambda>0$ is set free and used as a fitting parameter alongside with $\alpha$. The quartic twist theory is apparently more successful than the quadratic theory. \nigh{Quantitative estimates of the error are given for both curves in Appendix~\ref{sec:error}}.}
	\label{fig:R_vs_d}
\end{figure}
The former case  $(\lambda=0)$ corresponds to the classical quadratic theory and was already considered in \cite{collings:anchoring}:  the only fitting parameter is $\alpha$, and for $K_{22}\approx0.7\,\mathrm{pN}$ (value obtained from \cite{zhou:elasticity_2014} for DSCG at concentration $16.0\,\%\, \mathrm{(wt/wt)}$ and temperature $25^\circ\mathrm{C}$) the best fit delivers $W_0\approx0.37\mu\mathrm{J}/\mathrm{m}^2$, in complete agreement with \cite{collings:anchoring}. Alternatively, we can set $\lambda$ free and seek the best fit of the data by adjusting both $\alpha$ and $\lambda>0$: the best fit then delivers 
\begin{equation}
\label{eq:a_W_0}
a\approx1\,\mu\mathrm{m}, \quad W_0\approx1.55\,\mu\mathrm{J}/\mathrm{m}^2.
\end{equation}
Fig.~\ref{fig:R_vs_d} represents the results of both fits. It suggests that the quartic twist theory ($\lambda>0$) fits better the data than the classical quadratic theory ($\lambda=0$). This is confirmed quantitatively by the error estimate presented in Appendix~\ref{sec:error}. 

\nigh{
A similar experiment was conducted in \cite{peng:patterning} to measure $W_0$ for photoinduced planar anchoring of DSCG at concentration $14.0\,\% (\mathrm{wt/wt})$ and  temperature $25^\circ\mathrm{C}$ (substrates were coated with dye SD1 separated from the LCLC by a layer of RM257).

The scaled light intensities  $I_\perp^\mathrm{min}$ and $I_\perp^{\mathrm{max}}$ were measured for different values of the thickness $d$ in a wedge cell for $\lambda_0=633\, \mathrm{nm}$ and  $\Delta n=-0.018$. By using $\Omega$ as fitting parameter to best reproduce the experimental data via the theoretical functions in \eqref{eq:I_min_perp} and \eqref{eq:I_max_perp}, in \cite{peng:patterning} the classical quadratic theory enabled to estimate $W_0=1.7 \pm 0.2\, \mu\mathrm{J/m^2}$.

The experimental data taken from Fig.~2 of \cite{peng:patterning} are represented in  Fig.\ref{fig:I_vs_d}. Here,  with the aid of \eqref{eq:I_min_perp} and \eqref{eq:I_max_perp}, they will be processed by the same fitting strategy applied above to the other experiment, for both $\lambda=0$ and $\lambda>0$.
\begin{figure}[h] 
	\centering
	\includegraphics[width=0.4\linewidth]{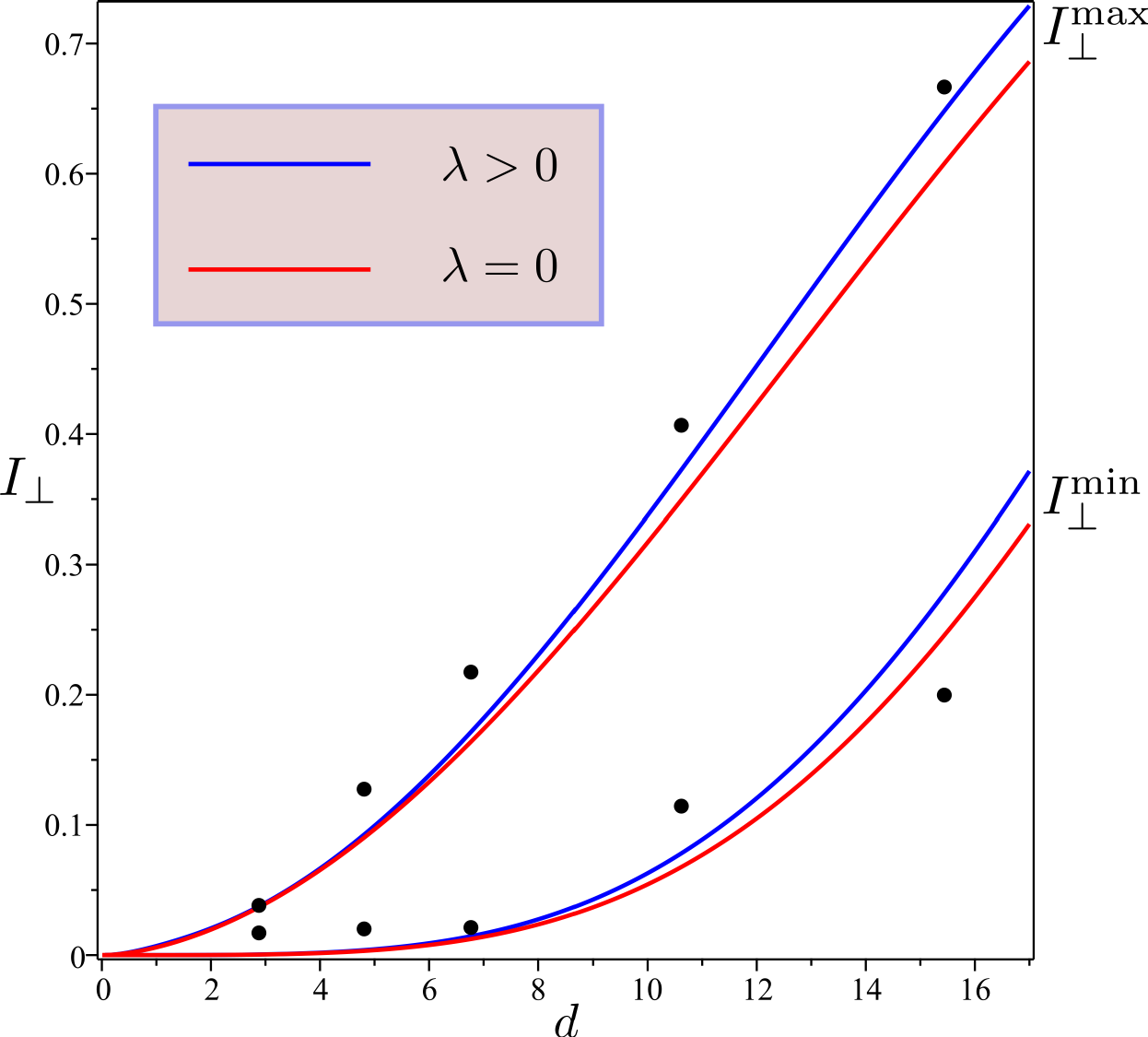}
	\caption{\nigh{Best combined fits of the data for the scaled light intensities  $I_\perp^\mathrm{min}$ and $I_\perp^{\mathrm{max}}$ in Fig.~2 of \cite{peng:patterning} with the theoretical functions given by \eqref{eq:I_min_perp} and \eqref{eq:I_max_perp} when the  total twist angle $\Omega$ is delivered by \eqref{eq:total_twist}.  Red curves (classical quadratic theory): $\lambda=0$ and $\alpha$ is the only fitting parameter. Blue curves (quartic twist theory): $\lambda>0$ is set free and used as a fitting parameter alongside with $\alpha$. Again, the quartic twist theory is more successful than the quadratic theory; further quantitative details are given in Appendix~\ref{sec:error}.}}
	\label{fig:I_vs_d}
\end{figure}
In the former case, the only fitting parameter is $\alpha$, and for $K_{22}\approx0.5\,\mathrm{pN}$ (value obtained from \cite{zhou:elasticity_2014} for DSCG at concentration $14.0\,\%\, \mathrm{(wt/wt)}$ and temperature $25^\circ\mathrm{C}$) the best fit delivers $W_0\approx1.51\,\mu\mathrm{J}/\mathrm{m}^2$, which falls within the error bar of the measurement of  \cite{peng:patterning}. The general best fit that utilizes  both $\alpha$ and $\lambda>0$ as adjusting parameters delivers 
\begin{equation}
	\label{eq:a_W_0_peng}
	a\approx4.8\,\mu\mathrm{m}, \quad W_0\approx1.65\,\mu\mathrm{J}/\mathrm{m}^2.
\end{equation}
Fig.~\ref{fig:I_vs_d} represents the results of both fits: once again, the quartic twist theory ($\lambda>0$) improves over the classical quadratic theory ($\lambda=0$). (Further quantitative details are  presented in Appendix~\ref{sec:error}.)

We finally comment on the difference (mainly in $a$) between the estimates in \eqref{eq:a_W_0} and \eqref{eq:a_W_0_peng}. As shown in \eqref{eq:R_def}, the data for the intensity ratio $R$ fitted in Fig.~\ref{fig:I_vs_d} ultimately contain only the data for the scaled intensity $I_\perp^\mathrm{max}$, whereas  the data for both $I_\perp^\mathrm{max}$ and $I_\perp^\mathrm{min}$ are jointly fitted in Fig.~\ref{fig:I_vs_d}. This inclines us to consider the estimate in \eqref{eq:a_W_0_peng} more reliable than the estimate in \eqref{eq:a_W_0}.
}

\section{Conclusion}\label{sec:conclusion}
A quartic twist theory for chromonic liquid crystals is applied to a $\pi/2$ twist cell. Like the classical quadratic theory, it predicts a linear twist within the cell, with a total twist angle related to both the anchoring strength $W_0$ at the bounding substrates and a new phenomenological length $a$.

Reinterpreting two recent experiments, we showed that the quartic twist theory affords a better representation of the data compared to the classical quadratic theory and estimates an anchoring strength $W_0$ approximately four times larger in one case, and nearly equal in the other. \nigh{The phenomenological length $a$ estimated here for DSCG turned out to be closer to the value estimated in \cite{paparini:elastic} for SSY ($a\approx6.4\,\mu\mathrm{m}$) than to that estimated in \cite{ciuchi:inversion} for SSY ($a$ ranging from $33\,\mu\mathrm{m}$ to $54\,\mu\mathrm{m}$ in different physical conditions). Such a wide range of values for $a$ is not yet understood, as we still lack a microscopic model to describe the origin and nature of this phenomenological parameter.}

We are aware that  a systematic collection of data would be required to achieve a sound experimental determination of both $W_0$  and $a$ for different materials and varying concentration and temperature. We hope that such an endeavour could be undertaken in the future; this paper has  only shown what is  perhaps one way to go.

\appendix
\section{Energy Comparison}\label{sec:comparison}
To make sure that the anchoring at the cell's plates is never broken, no matter how small $W_0$ is, we need compare the minimum total energy $\mathcal{F}$ computed on the linear twist solving \eqref{eq:g_equation}, with the energy of the broken anchoring in \eqref{eq:energy_c_0}.

To this end, we first note that by \eqref{eq:g_equation} the minimum of $\mathcal{F}$ can easily be expressed as a function $F$ of $\delta_0$,
\begin{equation}
	\label{eq:energy_twist}
	F(\delta_0):=\frac{1}{2}\left(\frac{\pi}{2}-2\delta_0\right)^2\left[1+\frac{\lambda^2}{2}\left(\frac{\pi}{2}-2\delta_0\right)^2\right]-\alpha\cos^2\delta_0.
\end{equation}
Since $\delta_0$ is determined numerically, and so we lack its explicit expression in terms of $\alpha$ and $\lambda$, we replace it with a test function $\delta_{\mathrm{test}}$ that reproduces  the same behaviour  as $\delta_0$ for $\alpha\to0$, $\alpha\to\infty$, and $\lambda\to\infty$,  and is chosen so that
\begin{equation}
	\label{eq:estimate_test}
	\Delta F_\mathrm{test}:=F(\delta_{\mathrm{test}})-\mathcal{F}[\vt_0]=F(\delta_{\mathrm{test}})+\frac{\alpha}{2}<0,
\end{equation}
where use has been made of \eqref{eq:energy_c_0}. This would suffice to prove that $F(\delta_0)<\mathcal{F}[\vt_0]$, as $F(\delta_0)\leqq F(\delta_{\mathrm{test}})$, being $F(\delta_0)$ the minimum of $\mathcal{F}$.

We took
\begin{equation}\label{eq:delta_test}
\delta_{\mathrm{test}}:=\frac{\pi}{2\left(\frac{\alpha}{1+\lambda^2}+2\right)},
\end{equation}
which, once inserted in \eqref{eq:estimate_test}, led us to show with little labour that, for any given $\lambda>0$, $\Delta F_\mathrm{test}$ is a decreasing function of $\alpha$, whose derivative vanishes only for $\alpha=0$, which possesses  the following asymptotic behaviours,
\begin{eqnarray}
\Delta F_\mathrm{test}&=&-\frac{\pi}{8(1+\lambda^2)}\left[1-\frac{\pi}{8(1+\lambda^2)}\right]\alpha^2+\mathcal{O}\left(\alpha^3\right),\label{eq:energy_test_alpha_0}\\
\Delta F_\mathrm{test}&=&-\frac{\alpha}{2}+\mathcal{O}\left(\frac{1}{\alpha}\right),\label{eq:energy_test_alpha_infinity}
\end{eqnarray}
for $\alpha\to0$ and $\alpha\to\infty$, respectively. This indeed proves our claim \eqref{eq:estimate_test}.

\section{Error Estimates}\label{sec:error}
Here we describe in some detail the strategy adopted to fit the data shown in Figs.~\ref{fig:R_vs_d} and \ref{fig:I_vs_d} of the main text.

First, we introduce a dimensionless parameter, $\lambda^\ast$, which is independent of the cell's thickness $d$,
\begin{equation}
\label{eq:lambda_lambda_star}
\lambda^\ast:=\alpha\lambda=\frac{aW_0}{K_{22}}.
\end{equation}
Then, with the aid of \eqref{eq:total_twist}, we rewrite \eqref{eq:g_equation} as an equation for $\Omega$ in the parameters $(\lambda,\lambda^\ast)$,
\begin{equation}
\label{eq:g_eta}
\Omega=\frac{1}{\lambda}\frac{6^{2/3}\left(\sqrt{(9\lambda^\ast\cos\Omega)^2+48}+9\lambda^\ast\cos\Omega\right)^{2/3}-12}{6^{4/3}\left(\sqrt{(9\lambda^\ast\cos\Omega)^2+48}+9\lambda^\ast\cos\Omega\right)^{1/3}}.
\end{equation}

\begin{figure}[h] 
	\centering
	\includegraphics[width=0.5\linewidth]{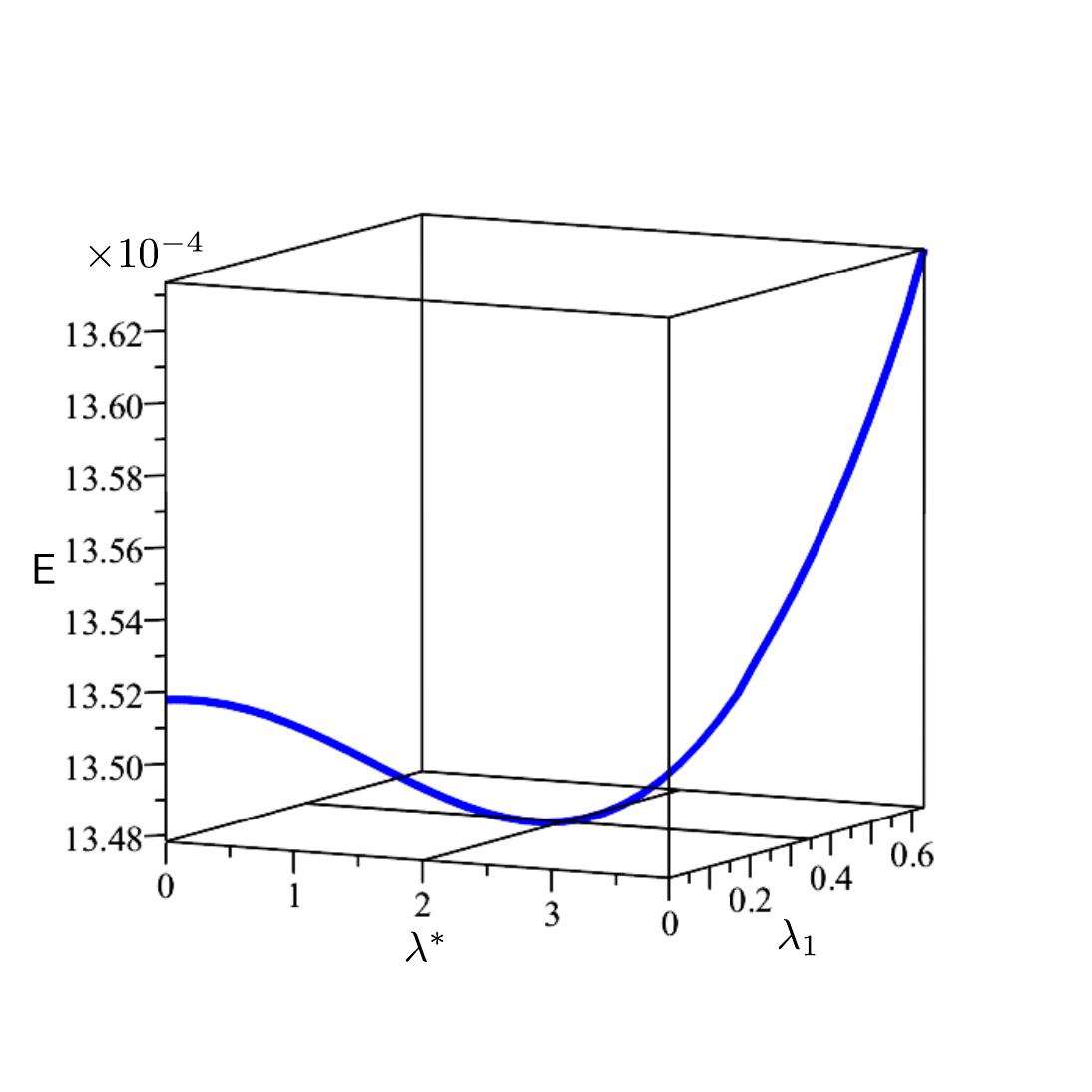}
	\caption{Three-dimensional graph illustrating the path $\lambda^\ast\mapsto\min_{\lambda_1}\mathsf{E}(\lambda_1,\lambda^\ast)$. In the limit as $\lambda^\ast\to0$, the minimum of $\mathsf{E}$ in $\lambda_1$ is attained at $\lambda_1=0$; its value equals the minimum of $\mathsf{E}_0$ in \eqref{eq:Error_def_T2}.}
	\label{fig:Emin3d}
\end{figure}

In the experiment performed in \cite{collings:anchoring}, eight values $d_i$ ($i=1,\dots,8$) of the thickness $d$ were selected, ranging from $d_1\approx2.9\,\mu\mathrm{m}$ to $d_8\approx14.7\,\mu\mathrm{m}$, and the corresponding values $R_i$ of  the intensity ratio $R$ were then measured.
We set 
\begin{equation}
	\label{eq:lambda1_def}
	\lambda_1:=\frac{a}{d_1}
\end{equation}
and regard this as a fitting parameter. Clearly, letting $\lambda_i:=a/d_i$, all these parameters are determined by $\lambda_1$ and the measured $d_i$'s,
\begin{equation}
	\label{eq:lambda_i}
	\lambda_i=\lambda_1\frac{d_1}{d_i}.
\end{equation}
For given $\lambda_1$ and $\lambda^\ast$, we denote $\Omega_i(\lambda_1,\lambda^\ast)$ the root of \eqref{eq:g_eta} with $\lambda=\lambda_i$ and compute the error
\begin{equation}
\label{eq:Error_def}
\mathsf{E}(\lambda_1,\lambda^\ast):=\frac{\sum_{i=1}^{8}\left[R(\Omega_i(\lambda_1,\lambda^\ast))-R_i\right]^2}{8},
\end{equation}
where $R(\Omega)$ is the function in \eqref{eq:R_def}. The function $\mathsf{E}$ has been studied numerically and found to have a single minimum in $\lambda_1$, for every given value of $\lambda^\ast>0$; the path of these constrained minima,  parametrized in $\lambda^\ast$, is shown in Fig.~\ref{fig:Emin3d}.
The absolute minimum of \nigh{$\mathsf{E}\approx13.48\times10^{-4}$} is attained at
\begin{equation}
\label{eq:lambda_1_star_val}
\lambda_1\doteq0.344, \quad \lambda^\ast\doteq2.01.
\end{equation}
For $K_{22}\approx0.7\mathrm{pN}$ and $d_1\approx2.9\,\mu\mathrm{m}$, from \eqref{eq:lambda_1_star_val} we estimate the optimal values for $a$ and $W_0$ in \eqref{eq:a_W_0}.

The graph of the curve that represents $R$ in Fig.~\ref{fig:R_vs_d} was obtained by plotting the function in \eqref{eq:R_def} for $\Omega$ given in terms of $d$ by the root of \eqref{eq:g_eta} for $\lambda^\ast$ as in \eqref{eq:lambda_1_star_val} and $\lambda=a/d$, where $a$ is the optimal value in \eqref{eq:a_W_0}. 

For $\lambda=0$, the error estimate goes along slightly different lines. By \eqref{eq:g_0} and \eqref{eq:total_twist}, \eqref{eq:g_eta} is now replaced by
\begin{equation}
	\label{eq:Omega_root_lambda_0}
\Omega=\frac{\alpha}{2}\cos\Omega.
\end{equation}	
We denote by $\Omega_i(\xi_1)$ the root of \eqref{eq:Omega_root_lambda_0} for $\alpha$ given by
\begin{equation}
	\label{eq:alpha_i}
	\alpha_i=\frac{1}{\xi_1}\frac{d_i}{d_1},
\end{equation}
where $\xi_1:=\el/d_1$. The latter  is treated as a fitting parameter featuring in the error
\begin{equation}
\label{eq:Error_def_T2}
\mathsf{E}_{0}(\xi_1):=\frac{\sum_{i=1}^{8}\left[R(\Omega_i(\xi_1)-R_i\right]^2}{8}.
\end{equation}
The minimum of $\mathsf{E}_0$ was found for $\xi_1\doteq0.655$, which for $K_{22}\approx0.7\,\mathrm{pN}$ corresponds to $\el\approx1.9\,\mu\mathrm{m}$ and $W_0\approx0.37\,\mu\mathrm{J/m}^2$, in complete agreement with \cite{collings:anchoring}. The minimum of $\mathsf{E}_{0}$ is approximately $13.52\times10^{-4}$, which equals the minimum of $\mathsf{E}$ in \eqref{eq:Error_def} for $\lambda_1=0$ (see Fig.~\ref{fig:Emin3d}). \nigh{Thus, the quartic theory affords an error that is estimated to be $3\,\%$ less than the error of the quadratic theory.} This proves that the data shown in Fig.~\ref{fig:R_vs_d} are indeed better approximated for $\lambda>0$.
The red curve that represents $R$ in Fig.~\ref{fig:R_vs_d} was obtained by plotting the function in \eqref{eq:R_def} for $\Omega$ given in terms of $d$ by the root of \eqref{eq:Omega_root_lambda_0} for $\alpha=d/\el$, where $\el$ is the optimal value just determined. 

\nigh{
Similarly, in the experiment performed in \cite{peng:patterning}, five values $d_i$ ($i=1,\dots,5$) of the thickness $d$ were selected, ranging from $d_1\approx2.9\,\mu\mathrm{m}$  to $d_5\approx15.5\,\mu\mathrm{m}$,  and the corresponding values of minimum and maximum relative intensities, $I_{\perp,i}^{\mathrm{min}}$ and $I_{\perp,i}^{\mathrm{max}}$, were measured.
	The error $\mathsf{E}$ is now more appropriately defined as
	\begin{equation}
		\label{eq:Error_def_bis}
		\mathsf{E}(\lambda_1,\lambda^\ast):=\frac{\sum_{i=1}^{5}\left(\left[I_\perp^\mathrm{max}(\Omega_i(\lambda_1,\lambda^\ast))-I_{\perp,i}^{\mathrm{max}}\right]^2+\left[I_{\perp}^\mathrm{min}(\Omega_i(\lambda_1,\lambda^\ast))-I_{\perp,i}^{\mathrm{min}}\right]^2\right)}{10}.
	\end{equation}
	The minimum in $\lambda_1$ of $\mathsf{E}$ for given $\lambda^\ast$ is shown in Fig.~\ref{fig:Emin3d_peng}.
	\begin{figure}[h] 
		\centering
		\includegraphics[width=0.5\linewidth]{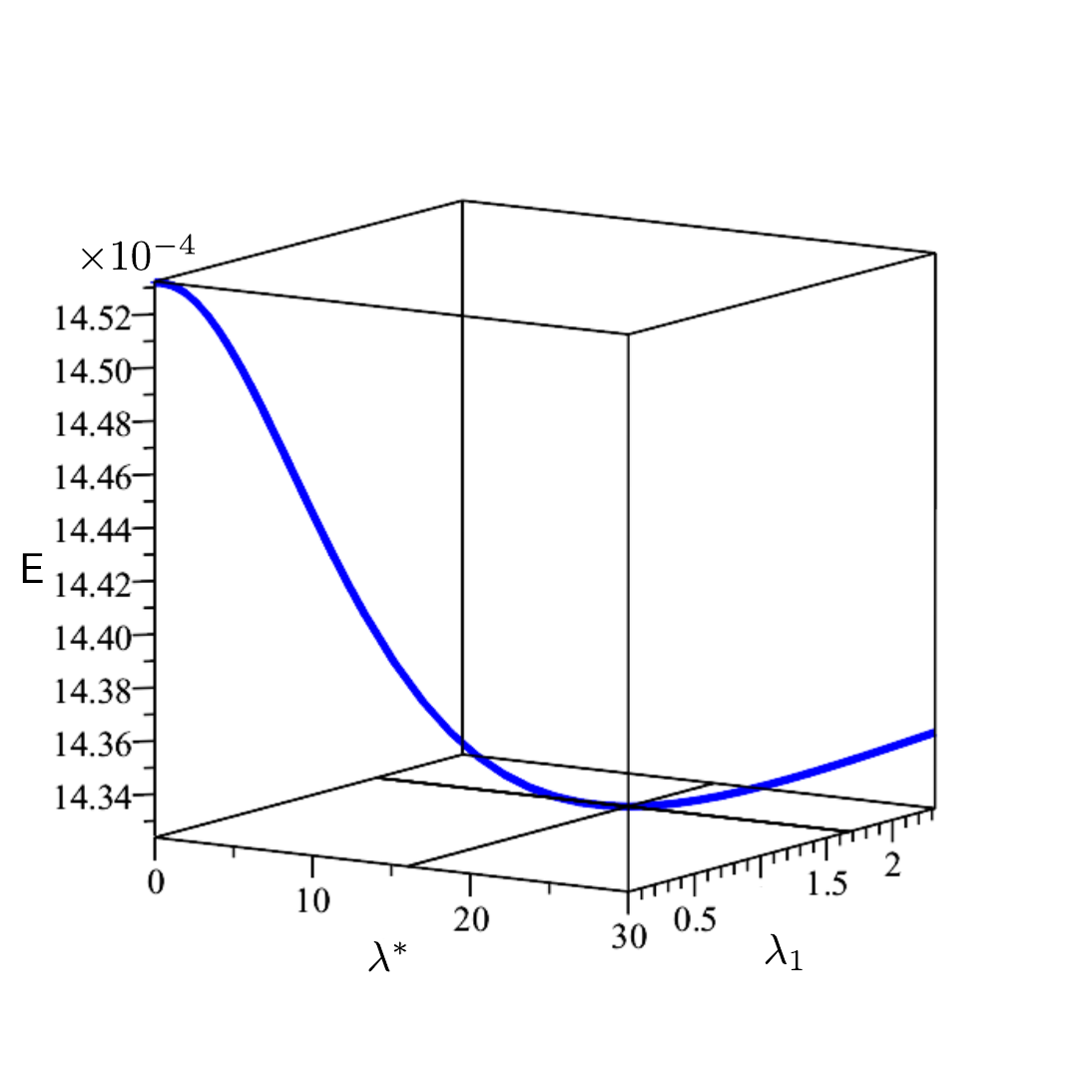}
		\caption{Three-dimensional graph illustrating the path $\lambda^\ast\mapsto\min_{\lambda_1}\mathsf{E}(\lambda_1,\lambda^\ast)$, very similar to the one in Fig.~\ref{fig:Emin3d}.}
		\label{fig:Emin3d_peng}
	\end{figure}
	The absolute minimum of $\mathsf{E}\approx 14.32\times10^{-4}$ is attained at
	\begin{equation}
		\label{eq:lambda_1_star_val_bis}
		\lambda_1\doteq1.67, \quad \lambda^\ast\doteq16.
	\end{equation}
	For $K_{22}\approx0.5\mathrm{pN}$ and $d_1\approx2.9\,\mu\mathrm{m}$, from \eqref{eq:lambda_1_star_val_bis} we estimate the optimal values for $a$ and $W_0$ in \eqref{eq:a_W_0_peng}.
	The graphs of the blue curves  in Fig.~\ref{fig:I_vs_d} were obtained precisely as that in Fig.~\ref{fig:R_vs_d}.
	
	For $\lambda=0$, the error is  given by
	\begin{equation}
		\label{eq:Error_def_T2_bis}
		\mathsf{E}_{0}(\xi_1):=\frac{\sum_{i=1}^{5}\left(\left[I_\perp^\mathrm{max}(\Omega_i(\xi_1))-I_{\perp,i}^{\mathrm{max}}\right]^2+\left[I_\perp^\mathrm{min}(\Omega_i(\xi_1))-I_{\perp,i}^{\mathrm{min}}\right]^2\right)}{10}.
	\end{equation}
	The minimum of $\mathsf{E}_0\approx14.53\times10^{-4}$ was found for $\xi_1\doteq0.11$, which for $K_{22}\approx0.5\,\mathrm{pN}$ corresponds to $\el\approx0.33\,\mu\mathrm{m}$ and $W_0\approx1.51\,\mu\mathrm{J/m}^2$. Thus, the error of the quartic theory is $1.4\,\%$ less than the error of the quadratic one.
	Finally, also the red curves in Fig.~\ref{fig:I_vs_d} were obtained precisely as that in Fig.~\ref{fig:R_vs_d}.
}
\begin{acknowledgements}
We are grateful to an anonymous Reviewer of an earlier version of this paper for their suggestions, which have improved our work.
\end{acknowledgements}
%\bibliographystyle{vancouver}
%\bibliography{Chromonics}
%\input{Collings_twist_rev_unmarked.bbl}

\end{document}